# DICE Fault Injection Tool

Craig Sheridan
Flexiant Limited
Geddes House
Livingston, Scotland
+44 (0)1506606009
csheridan@flexiant.com

Darren Whigham
Flexiant Limited
Geddes House
Livingston, Scotland
+44 (0)1506606000
dwhigham@flexiant.com

Matej Artač
XLAB d.o.o.
Pot za Brdom 100
SI-1000 Ljubljana, Slovenia
+386 (0)1 244 77 53
matej.artac@xlab.si

## ABSTRACT
In this paper, we describe the motivation, innovation, design, running example and future development of a Fault Inject Tool (FIT). This tool enables controlled causing of cloud platform issues such as resource stress and service or VM outages, the purpose being to observe the subsequent effect on deployed applications. It is being designed for use in a DevOps workflow for tighter correlation between application design and cloud operation, although not limited to this usage, and helps improve resiliency for data intensive applications by bringing together fault tolerance, stress testing and benchmarking in a single tool.

## 1. INTRODUCTION
Operating data intensive applications unavoidably involves dealing with various failures. Industry has long recognised instead of designing software to never fail, a more viable strategy is to make it reliable and resilient to failure[1]. To test applications during development, the FIT will be invaluable towards quality-driven application development in a DevOps environment.

The DICE [2] FIT allows the tool user to generate faults either at VM or cloud level. The purpose is to allow cloud platform owners a means to test the resiliency of a cloud installation as an application target. With this approach, the designers can use robust testing, showing where to harden the application before it reaches a commercial environment and allows a user/application owner to test and understand their application design/ deployment in the event of a cloud failure or outage. Thus allowing for the mitigation of risk in advance of a cloud based deployment.

## 2. MOTIVATION
Projected growth in the big data market shows Data Centre owners, cloud service providers and application owners with their demanding data intensive requirements are all potential beneficiaries of the FIT.

One key aspect that can affect the application QoS is the resilience of the underlying infrastructure. With FIT, the Data Centre owners can gauge the stress levels on different parts of the infrastructure, and thus offer advice to their customers, address bottlenecks or adapt the pricing of various levels of assurances.

To developers, FIT provides a tool for evaluating application resilience and dependability, which can only be demonstrated at runtime through forcing parts of the application to fail. By designing FIT to be lightweight and versatile command line tool it's trivial to use it during Continuous Integration or within another tool for running complex failure scenarios. Used in conjunction with other tools not within the scope of this paper, FIT could monitor and evaluate the effect faults have on an application and provide feedback to developers on design.

### 2.1 Innovation
Other fault injection tools exist in the market, such as ChaosMonkey[3] created by Netflix. These tools are either platform specific or limited in functionality. For instance, ChaosMonkey only accesses AWS and terminates VMs. The DICE FIT will address the need to generate various cloud agnostic faults at the VM Admin and Cloud Admin levels.

This range of functionalities allows greater flexibility and the ability to generate multiple faults. In addition when compared to other fault injection tools, the FIT is lightweight and only installs the required tools and components on the target VMs.

Other alternatives such as Cocoma[4] have been developed but not supported and suffer from limitations such as high resource usage, complex configuration as well as limited extensibility. FIT has been designed with future extensibility in mind, carefully considering the needs and challenges of cloud service providers such as the scalability and resiliency of the cloud consumption marketplace. The aim is to provide an important component for quality-driven development of applications in cloud software stacks. This will lead to more reliable Big Data and IoT solutions through early discovery and ability to address stability issues.

## 3. DESIGN
FIT allows VM Admins, application owners and Cloud Admins to generate various faults. It runs independently and externally to any target environment. Figure 1 illustrates the architecture.

To access the VM level and issue commands the DICE FIT uses SSH to connect to the Virtual Machines and issue the commands. By using JSCH, the tool is able to connect to any VM that has SSH enabled and issue commands as a pre-defined user. This allows greater flexibility of commands as well as the installation of tools and dependences.

The DICE FIT will be released under the permissive APACHE LICENSE 2.0 and supports the OS configurations of Ubuntu -

Tested with Ubuntu versions 14.0, 15.10 and Centos (With set Repo configured & wget installed) - Tested with Centos version 7. The faults to be supported by the FIT are described in Table 1.

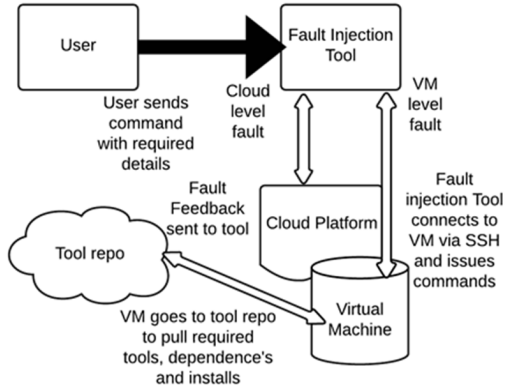

**Figure 1: Fault Injection Tool architecture**

**Table 1. FIT Operations**

| Access Level | Operation |
|---|---|
| Cloud Admin | Shutdown node |
| | High CPU for Node |
| | High Memory usage for Node |
| | High Bandwidth usage. |
| VM Admin | Shutdown random VM |
| | High CPU for VM |
| | High Memory usage for VM |
| | Block VM external access |
| | High Bandwidth usage. |
| | High Disk I/O usage |
| | Stop service running on VM |
| | Shutdown random VM from whitelist provided by user |
| | YCSB on VM running MongoDB to begin workload test |
| | Run JMeter plan |

## 3.1 Running Example on the Real System

The FIT uses a command line interface with a number of command line switches and parameters[5], letting the users select a specific fault to be injected and the parameters of the fault such as the amount of RAM to use or which services to stop. This is an example of a command line call to connect to a node using SSH credentials and stress the memory with 2 GB:

```
--stressmem 2 2048m ubuntu@109.231.126.101 -no home/ubuntu/SSHKEYS/VMkey.key
```

The tool first connected via SSH to the VM and determined the OS version (Ubuntu in our case). It then looked for the memory stress tool suitable for Ubuntu, in this case Memtester[6]. If the tool is not found first the DICE FIT installs the tool along with dependencies. Finally, the FIT called Memtester to saturate memory in the target node. Figure 2 shows the result of this, as detected by a monitoring tool, showing nearly all of the 2GB of RAM available to the VM being saturated.

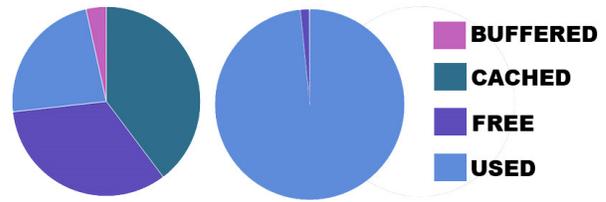

**Figure 2: Memory available on the target node before (left) and during FIT's invocation (right)**

## 4. CONCLUSION

In this paper, we presented the DICE Fault Injection tool aimed at developers using a quality driven DevOps approach to be able to test their applications under the conditions of failing parts of the infrastructure or misbehaving services. The tool is easy to use and integrate in the Continuous Deployment workflow. In the current implementation of the tool, we show support of VM level faults.

Future work will focus on difficulties and possibilities of extensibility by external users, investigating limitations, consider different topologies, operating systems and vendor agnostic cloud provider infrastructure as well as evaluating the overhead of operation. Containerised environments will also be considered as future FIT targets to help understand the effect on microservices when injecting faults to the underlying host as well as the integrity of the containerised deployment. Further, the CACTOS[7] project will expand the tool functionality by initiating a specific application level fault to trigger optimisation algorithms. In the longer term other target cloud APIs could be added to the FIT as well as investigating related academic work.

## 5. ACKNOWLEDGMENTS

The authors' work is partially supported by the European Commission grant no. 644869 (H2020 - Call 1), DICE.